\documentclass[preprint,showpacs,preprintnumbers,amsmath,amssymb,showkeys]{revtex4}
\usepackage{graphicx}
\usepackage{hhline}
\usepackage{dcolumn}
\usepackage{bm}

\begin{document}

\title{Ferromagnetic Ising spin systems on the growing random tree}

\author{Takehisa Hasegawa} 
\email{hase@statphys.sci.hokudai.ac.jp}
\author{Koji Nemoto}
\affiliation{%
Division of Physics, 
Hokkaido University, Sapporo, Japan.
}%

\begin{abstract}
We analyze the ferromagnetic Ising model on a scale-free tree; 
the growing random network model with the linear attachment kernel $A_k=k+\alpha$ 
introduced by [Krapivsky et al.: Phys. Rev. Lett. {\bf 85} (2000) 4629-4632]. 
We derive an estimate of the divergent temperature $T_s$ below which 
the zero-field susceptibility of the system diverges.
Our result shows that 
$T_s$ is related to $\alpha$ as $\tanh(J/T_s)=\alpha/[2(\alpha+1)]$, 
where $J$ is the ferromagnetic interaction.
An analysis of exactly solvable limit for the model and numerical calculation support the 
validity of this estimate.
\end{abstract}
\pacs{89.75.Fb, 05.50.+q, 89.75.Hc}

\keywords{Ising model, critical phenomena, zero-field susceptibility, complex network, growing tree, scale-free degree distribution}
\maketitle

\section{Introduction}

Many real-world systems range from 
the structure of Internet or WWW to 
social relationship between human society, 
or prey-predator relationship in food webs
are described topologically as scale-free networks (SFNs) \cite{BA,AB,NewRev}.
In a SFN, the degree distribution
$P(k)$, where degree $k$ is the number of edges connected
to a node,
has a power-law decay
$P(k) \propto k^{-\gamma}$. 
The degree exponent  $\gamma$ 
takes $2 < \gamma < 3$ in many real networks \cite{Get}.
SFN studies 
have been carried out actively in recent years, 
including 
%
various processes taking thereon, 
e.g., network failure, spread of infections, 
or interacting systems, which attract  
numerous applications and fundamental interests 
about critical phenomena \cite{AB,NewRev,DorogoRev}. 
Among them,
the ferromagnetic Ising model on SFNs is a basic prototype to understand 
how network topology influences cooperative behaviors on SFNs. 
Indeed, previous studies 
have demonstrated the extreme influence of the network topology
by both analytical \cite{Bi,Let,IT,Det02,SH,HB} and numerical methods \cite{AHS,H,ZL}.
For example, Dorogovtsev et al. \cite{Det02} analyzed 
the Ising model on an uncorrelated SFN model
with a power-law degree distribution by the Bethe approach 
to show that
the phase transition exists at a finite temperature for $\gamma>3$, 
while the system remains in the ferromagnetic phase at any finite temperature for $\gamma \le 3$, 
and its critical exponents vary depending on $\gamma$.

In the previous approaches,
rather dense part of a SFN has been analyzed in details, 
while our knowledge of how the spins on {\it leaves} (nodes whose degree is one) behave is still missing.
In practice, many real SFNs have a number of leaves. 
How does such network topology influence the critical behaviors?
This paper focuses on a simple case; 
the Ising model on a tree-like SFN which includes many leaves, but has no cyclic path. 
We note that the ferromagnetic Ising model on trees behaves quite differently from 
that on regular lattices; 
the spin system on a tree has no spontaneous magnetization at any finite temperature, 
while its zero-field susceptibility remains to diverge below a certain temperature $T_s$ 
(we call it the {\it divergent temperature}) \cite{E,MHZ,M74,HT,F,Met,Set98,HN}.
In other words, the leaf-spins are extremely sensitive to the external field below $T_s$.
Our aim is to clarify the relation between the divergent temperature $T_s$ and the degree distribution $P(k)$.

In this paper, we analyze the Ising model on the {\it growing random network} (GN) {\it model} 
introduced by Krapivsky and coworkers \cite{KR00}.
The GN model probabilistically generates a sampled tree ${\rm T}_N$ with $N$ nodes as follows.
One starts with one isolated node. 
At each time step, 
a new node is added and links to a preexisting node.
The probability that the new node 
links to a node with degree $k$ is proportional to 
{\it the attachment kernel} $A_k$.
In this paper, we focus on the linear attachment kernel $A_k=k+\alpha$ ($\alpha>-1$).
The degree distribution of a resulting tree is determined by the offset $\alpha$ 
in the attachment kernel.
For the infinite offset $\alpha \to \infty$, 
the degree distribution reduces to the exponential form $P(k)=2^{-k}$.
For a finite offset $\alpha$, the degree distribution satisfies 
a power-law decay $P(k) \propto k^{-\gamma}$, 
where the degree exponent is $\gamma =3+\alpha$ \cite{KR00}.
As the offset $\alpha$ is smaller, the resulting degree distribution $P(k)$ is more fat-tailed.
Particularly, the GN model with the no offset $\alpha=0$ 
corresponds to so-called Barab\'asi-Albert tree
\cite{BR04,FVK06,SAK02,MetDis}, 
which has the degree exponent $\gamma =3$.
We investigate how the divergent temperature $T_s$ is related to $\alpha$ 
to show the extreme sensitivity of trees with fat-tailed degree distribution ($2<\gamma \le 3$).

This paper is organized as follows.
In section \ref{model}, we introduce the Ising model on the GN model.
In section \ref{infinite}, we derive the exact expression for the system susceptibility and the divergent temperature
for the infinite offset case $\alpha \to \infty$. 
In section \ref{finite}, we give an estimate of $T_s$ for the general offset case. 
In section \ref{Numcal}, we show our numerical results to support the validity of our estimate, and  
suggest that for the no offset case, $T_s$ diverges and an unusual scaling exists. 
Section \ref{summary} is devoted to summary.

\section{Model \label{model}}

In this section, we introduce our model; 
the Ising model on the GN model.
The Hamiltonian is as follows:
\begin{equation}
\mathcal{H}=-J \sum_{\langle i j \rangle} S_iS_j -h \sum_i S_i ,
\end{equation}
where $J (>0)$ is the ferromagnetic interaction, 
$h$ is the external magnetic field,
and $S_i (=\pm 1)$ is the Ising spin variable on the node $i$.
The first sum is over all edges of a network,
and the second one is over all nodes.
In the following sections, we calculate the zero-field susceptibility 
of this model.
The zero-field susceptibility is expressed in terms of 
the spin-spin correlation functions as 
$\chi_{\rm sys}=\frac{1}{N}\sum_{i,j=1}^N \langle S_iS_j \rangle$, 
where the angular bracket denotes the usual thermal average.
For trees, 
the correlation function between two Ising spins $S_i$ and $S_j$ on a sampled tree ${\rm T}_N$
is given as \cite{F}
\begin{equation}
\langle S_i S_j \rangle =t^{d_{i,j}({\rm T}_N)}, \label{Falktheorem}
\end{equation}
where $t=\tanh(\beta J)$, $\beta =1/T$, $T$ being the temperature, and 
$d_{i,j}({\rm T}_N)$ is the path length between the node $i$ and $j$ on the tree ${\rm T}_N$.
Accordingly,
the {\it one-spin} susceptibility $\chi_i ({\rm T}_N)$ of a spin on the node $i$ 
of a sampled tree ${\rm T}_N$ is 
\begin{equation}
\chi_i ({\rm T}_N) =\beta \sum_{j=1}^N t^{d_{i,j}({\rm T}_N)},
\end{equation}
and the system susceptibility 
\begin{equation}
\chi_{\rm sys}({\rm T}_N)=\frac{1}{N}\sum_{i=1}^N \chi_i({\rm T}_N)
=\frac{\beta}{N}\sum_{i,j=1}^N t^{d_{i, j}({\rm T}_N)}.
\end{equation}
Note that the system susceptibility is related with the so called {\it average correlation volume} 
$\xi_V$ \cite{DorogoRev}: $\xi_V=T\chi_{\rm sys}$. 

\section{infinite offset case \label{infinite}}

In this section, we consider the GN model with the infinite offset $\alpha \to \infty$.
We derive the exact form for the mean system susceptibility 
$\overline{\chi_{\rm sys}}(N,T)$ and the divergent temperature $T_s$.
Here $\overline{A}=\sum_{{\rm T}_N}P({\rm T}_N)A({\rm T}_N)$, and 
$P({\rm T}_N)$ is the normalized probability of a tree ${\rm T}_N$.
Suppose that T$_{N+1}$ is created by attaching the $(N+1)$-th node to the $n$-th node of a preexisting tree T$_N$.
Then the distance from the new node to all others is given as 
$d_{i,N+1}({\rm T}_{N+1})=d_{N+1,i}({\rm T}_{N+1})=1+d_{n,i}({\rm T}_N)$ for $1 \le i \le N$ \cite{MetDis}.
The diagonal element is zero: $d_{N+1,N+1}({\rm T}_{N+1})=0$.
Note that for trees,  
a path between each two nodes is unique,
 so the matrix elements do not change their values once formed during the growth process.
So we obtain the recursion relation for 
the averaged total susceptibility 
$\beta V_N = N \overline{\chi_{\rm sys}}(N,T)$ as follows:   
\begin{eqnarray}
V_{N+1} &=& \sum_{{\rm T}_{N+1}} P({\rm T}_{N+1}) \sum_{i, j =1}^{N+1}t^{d_{i,j} ({\rm T}_{N+1})}  \nonumber \\
&=& 
\sum_{{\rm T}_{N+1}} P({\rm T}_{N+1}) \sum_{i, j =1}^{N}t^{d_{i,j} ({\rm T}_{N+1})}+ 
2 \sum_{{\rm T}_{N+1}} P({\rm T}_{N+1}) \sum_{i =1}^{N}t^{d_{i,N+1} ({\rm T}_{N+1})}+1. \label{BATparts1}
\end{eqnarray}
The first term of the r.h.s. is 
\begin{equation}
\sum_{{\rm T}_{N+1}} P({\rm T}_{N+1}) \sum_{i, j =1}^{N}t^{d_{i,j} ({\rm T}_{N+1})}=\sum_{{\rm T}_{N}} P({\rm T}_{N}) \sum_{i, j =1}^{N}t^{d_{i,j} ({\rm T}_{N})} =V_N, \label{BATparts2}
\end{equation}
and the second term of the r.h.s. is 
\begin{eqnarray}
\sum_{{\rm T}_{N+1}} P({\rm T}_{N+1}) \sum_{i=1}^{N}t^{d_{i, N+1} ({\rm T}_{N+1})} &=& 
\sum_{{\rm T}_{N}} P({\rm T}_{N}) \sum_{n=1}^N P(n|{\rm T}_N) \sum_{i =1}^{N}t^{1+ d_{i,n} ({\rm T}_{N})} .  \label{BATparts3}
\end{eqnarray}
Here we use $P({\rm T}_{N+1}) = P({\rm T}_N)P(n|{\rm T}_N)$, 
where $P(n|{\rm T}_N)$ is the conditional probability that
the newly-added node links to a preexisting node labeled as $n$ on the tree ${\rm T}_N$.
Combining Eqs.(\ref{BATparts2}) and (\ref{BATparts3}) with Eq.(\ref{BATparts1})
gives the evolution of $V_N$ as
\begin{equation}
V_{N+1}=V_{N}+1+ 2t \sum_{{\rm T}_{N}} P({\rm T}_{N}) \sum_{n=1}^N P(n|{\rm T}_N) \sum_{i =1}^{N}t^{d_{i,n} ({\rm T}_{N})}. \label{BATfrec}
\end{equation}
For the general offset case, 
it is hard to solve $V_N$ expicitly
since the probability $P(n|{\rm T}_N)$ is proportional to the kernel $A_k$. 
Fortunately, the infinite offset case is within reaching distance.
In this case, 
the conditional probability is independent of which 
node is attached: $P(n|{\rm T}_N)=\frac{1}{N}$ for any $n$. 
Thus Eq.(\ref{BATfrec}) is evaluated as 
\begin{eqnarray}
V_{N+1} = V_{N}+1+\frac{2t}{N} \sum_{{\rm T}_{N}} P({\rm T}_{N}) \sum_{n=1}^N \sum_{i =1}^{N}t^{d_{i,n} ({\rm T}_{N})} 
= 1+ \Big(1+ \frac{2t}{N} \Big) V_{N}. \label{frecexpotree}
\end{eqnarray}
This recursion equation is solved explicitly to obtain the mean system susceptibility as
\begin{equation}
\frac{\overline{\chi_{\rm sys}}(N,T) }{\beta} =\frac{V_N}{N}= 1+t+ 2t \sum_{m=2}^{N-1} \frac{1}{m(m+1)}
 \prod_{k=1}^{m-1} \Big(1+ \frac{2 t}{k} \Big). \label{totalchiform}
\end{equation}
By evaluating the temperature below which 
the system susceptibility (\ref{totalchiform}) diverges,
we find that 
the divergent temperature is given as 
(see appendix \ref{derdivExpoTree})
\begin{equation}
\tanh(J/T_s)=\frac{1}{2}. \label{infiniteDivT}
\end{equation}
Moreover, expanding Eq.(\ref{totalchiform}) around the divergent temperature, 
we obtain the finite size scaling form for the infinite offset case:
\begin{equation}
\overline{\chi_{\rm sys}}(N,T) \simeq (\log N) f[(T-T_s)\log N], \label{infiniteFSS}
\end{equation}
where $f(x)$ is a scaling function in this case.

\section{general offset case \label{finite}}

In this section, we proceed to the general offset case. 
We give an estimate of $T_s$ by calculating a lower bound of the system susceptibility. 
First, we can calculate approximately the one-spin susceptibility 
of the initial node $\overline{\chi_1}$.
Our calculation is based on a mean field approach by Szabo et al.\cite{SAK02}; 
the original stochastic model is approximated by a uniform branching tree 
where every node on any level has the same number of offsprings.
Let $n_N^{(l)}$ denote the number of nodes at the $l$-th level, 
which means the distance from the initial node is $l$, 
on the tree with $N$ nodes.
When the new node is added to the tree with $N$ nodes, 
the probability that the new node links to any node at the $l$-th level is 
\begin{eqnarray}
\frac{n_N^{(l)}+n_N^{(l-1)}+\alpha n_N^{(l)}}{(2+\alpha)N-1}. \nonumber
\end{eqnarray}
Here the new node is stationed at the $(l+1)$-th level, 
so we obtain 
\begin{equation}
n_{N+1}^{(l+1)}=n_N^{(l+1)}+\frac{ c_1 n_N^{(l)}+n_N^{(l+1)}}{ c_2 N-1} \quad (l \ge 1),  \label{nNlrec}
\end{equation}
where 
$c_1=1+\alpha$, $c_2=2+\alpha$, 
and the initial condition is $n_N^{(0)}=1$ for all $N$.
Now we introduce the generating function 
\begin{equation}
G_N(t)=\sum_{l=0}^\infty n_N^{(l)}t^l.
\end{equation}
Note that $G_N(t)$ is related to $\overline{\chi_1}$ 
as $\overline{\chi_1} =\beta G_N(t)$.
From Eq.(\ref{nNlrec}), we obtain the recursion relation for the generating function as follows: 
\begin{equation}
(c_2 N-1)G_{N+1}(t)=(c_2 N+c_1 t)G_N(t) -1.
\end{equation}
It is easily solved that 
\begin{eqnarray}
G_N  
= 1+t \frac{\Gamma(2-{c_2}^{-1})}{\Gamma(1+{c_2}^{-1} c_1 t)} 
\sum_{M=1}^{N-1} \frac{\Gamma(M+{c_2}^{-1} c_1 t)}{\Gamma(M+1-{c_2}^{-1})}.
\end{eqnarray}
For $N \gg 1$, the summation of the second term can be approximated as 
\begin{eqnarray}
\sum_{M=1}^{N-1} \frac{\Gamma(M+{c_2}^{-1} c_1 t)}{\Gamma(M+1-{c_2}^{-1})} 
\simeq \sum_{M=1}^{N-1} M^{\frac{c_1 t +1}{c_2}-1} 
\simeq N^{\frac{c_1 t +1}{c_2}}, 
\end{eqnarray}
so that 
\begin{eqnarray}
G_N(t) \simeq N^{\frac{c_1 t +1}{c_2}}.
\end{eqnarray}
Thus we obtain the one-spin susceptibility of the initial node as 
\begin{equation}
\overline{\chi_1} (N,T) =\beta G_N(t) \simeq \beta N^{\frac{1+(1+\alpha)t}{2+\alpha}},
\end{equation}
which diverges for any $T$ and any $\alpha (>-1)$.

Now we evaluate a lower bound of the system susceptibility.
We consider a subtree which consists of a node at the $s$-th level and its descendents. 
The number of node at the $(s+l)$-th level is given as $n_N^{(s+l)}/n_N^{(s)}$.
Among the total susceptibility of the subtree, 
the contribution from the node-pairs whose paths go through the level $s$ is 
$G_{N,s}^2- [n_N^{(s+l)}/n_N^{(s)}] t^2 G_{N,s+1}^2$. 
Here
\begin{equation}
G_{N,s} =\sum_{l=0}^\infty \frac{n_N^{(s+l)}}{n_N^{(s)}} t^l,  
\end{equation}
corresponds to the one-spin susceptibility of the node at the $s$-th level.
The total susceptibility of the whole tree is evaluated as 
\begin{eqnarray}
N T \overline{\chi_{\rm sys}}(N,T) &=& 
\sum_{s=0}^{\infty} \Big( G_{N,s}^2-t^2 G_{N,s+1}^2 \frac{n_N^{(s+l)}}{n_N^{(s)}} \Big) n_N^{(s)} \nonumber \\
&=&  n_N^{(0)} G_{N,0}^2 +(1-t^2) \sum_{s=0}^{\infty} n_N^{(s+l)} G_{N,s+1}^2.
\end{eqnarray}
The second term is non-negative for $0 \le t \le 1$, so we obtain a lower bound of the system susceptibility as
\begin{eqnarray}
T \overline{\chi_{\rm sys}}(N,T)   
\ge \frac{1}{N} n_N^{(0)} G_{N,0}^2 =\frac{1}{N} G_N^2 \simeq N^{2\frac{1+c_1 t}{c_2}-1}. 
\end{eqnarray}
Note that the exponent includes $t$. 
By evaluating where this bound diverges, we obtain an estimate of $T_s$: 
\begin{equation}
\tanh(J/T_s)=\frac{\alpha}{2(\alpha+1)}, \label{estimate}
\end{equation}
which reduces the exact solution (\ref{infiniteDivT}) 
for the infinite offset $\alpha \to \infty$.
This relation indicates that 
as the offset is smaller, the divergent temperature is higher.
Particularly, we immediately find that $T_s$ diverges for $-1 <\alpha \le 0$. 

\section{Numerical calculations \label{Numcal}}

\begin{figure}
\begin{center}
\includegraphics[width=10cm]{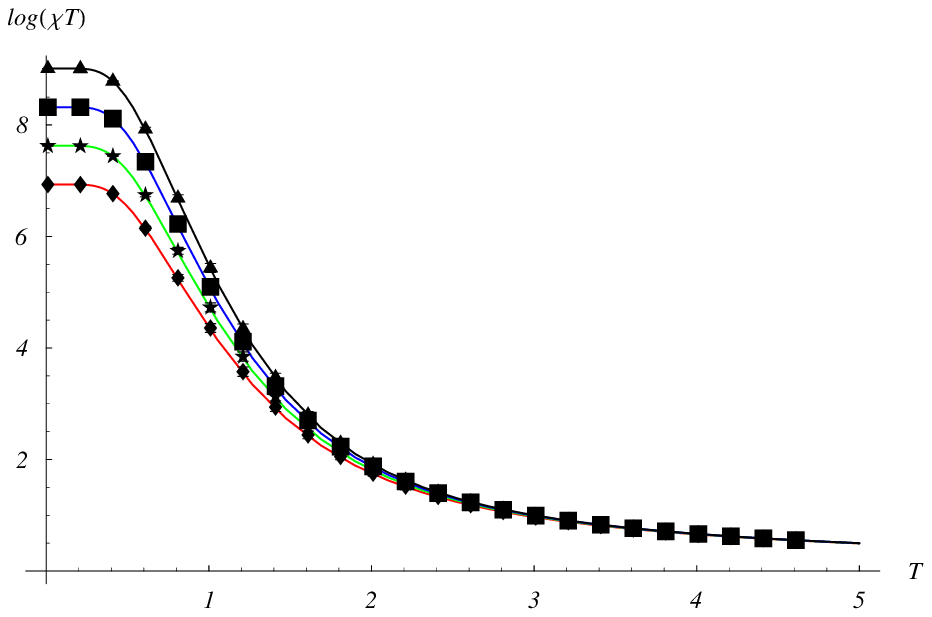}
\end{center}
\caption{
The system susceptibility for the infinite offset by 
the analytical results (\ref{totalchiform}) (lines) and the numerical ones (points).
The number of nodes are taken $N=2^{10}$(red), $2^{11}$(green),  $2^{12}$(blue), and $2^{13}$(black), respectively. 
The average is taken over $100$ samples.
}
\label{comparisonplot}
\end{figure}

\begin{figure}
\begin{center}
\includegraphics[width=10cm]{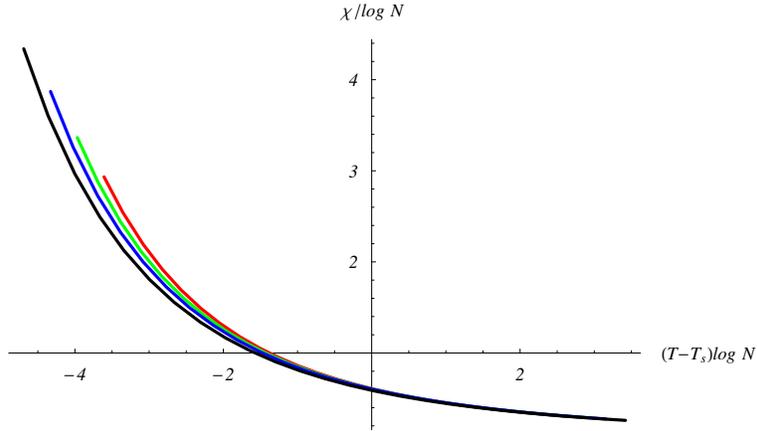}
\end{center}
\caption{
Finite size scaling (\ref{infiniteFSS}) 
of the system succeptibility around $T_s$ for the infinite offset. 
The number of nodes are taken $N=2^{10}$(red), $2^{11}$(green),  $2^{12}$(blue), and $2^{13}$(black), respectively. 
The average is taken over $100$ samples. 
}
\label{finitesizescalingoffsetinfinity}
\end{figure}

In this section, we calculate the zero-field susceptibilities numerically 
for the GN models with several values of offset $\alpha$. 
We generate trees for a given offset to 
calculate the susceptibilities by using Eq.(\ref{Falktheorem}). 
First, we show the results for the infinite offset case $\alpha \to \infty$.
Figure \ref{comparisonplot} compares the numerical result 
for the system susceptibility with the analytical forms (\ref{totalchiform}).
For convenience, we set $J=1$.
We find that the analytical forms agree well with the numerical ones.
Figure \ref{finitesizescalingoffsetinfinity} plots the finite size scaling around the divergent temperature.
The number of nodes is taken from $2^{10}$ to $2^{13}$.
The system susceptibilities are averaged over $100$ samples.
We find that the scaling works quite well.

\begin{figure}
\begin{center}
\includegraphics[width=11cm]{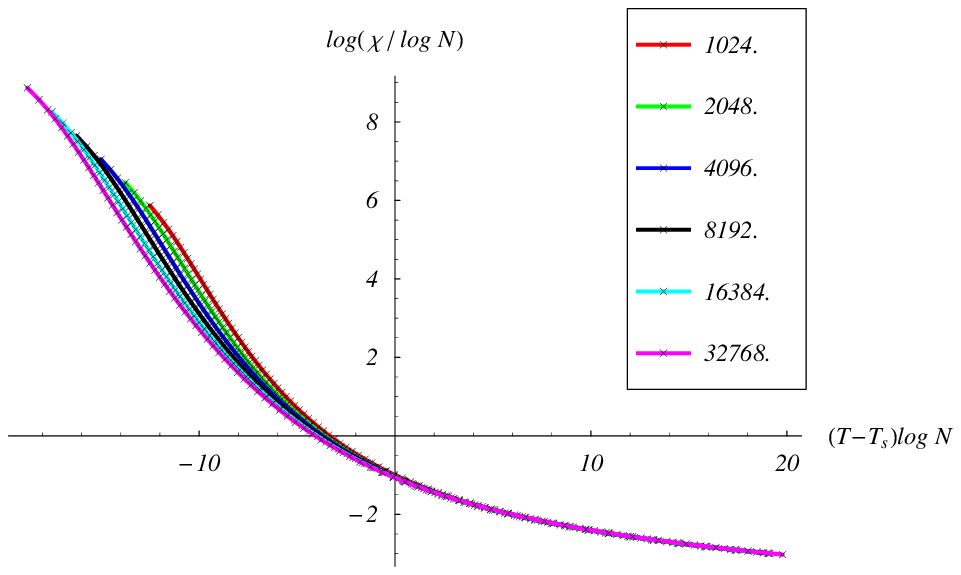}
\includegraphics[width=11cm]{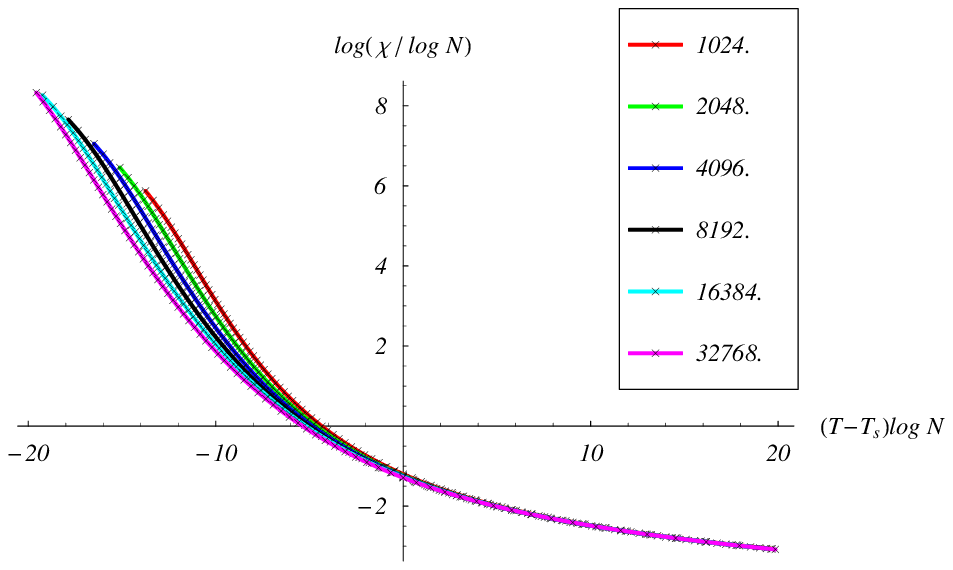}
\includegraphics[width=11cm]{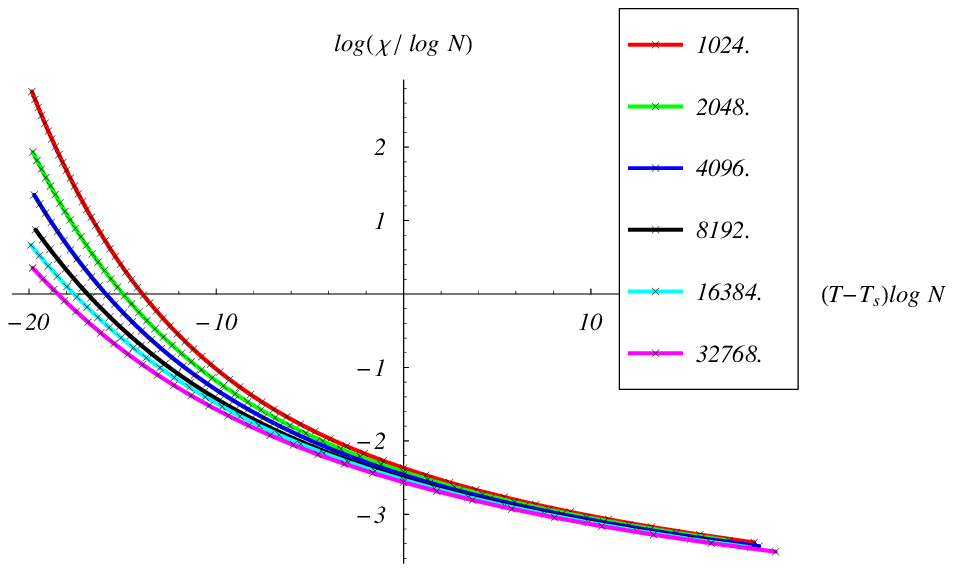}
\end{center}
\caption{
Finite size scaling 
of the system succeptibility around our estimate 
$T_s$ (\ref{estimate}) 
for the offset $\alpha=6$(top), 4(center), and 1(bottom). 
The number of nodes are taken from $N=2^{10}$ to $N=2^{15}$. 
The average is taken over $100$ samples.
}
\label{finitesizescalingoffset}
\end{figure}

Next, we turn to the finite offset case.
Figures \ref{finitesizescalingoffset} plots the finite size scaling around our estimate $T_s$ 
for the mean system susceptibility with the offset $6$, $4$, and $1$. 
As a result, we find that 
our finite size scaling similar to that for the infinite offset (\ref{infiniteFSS})
is quite well fitted as long as an offset is not small.
These results support that our estimate gives the exact divergent temperature. 
On the other hand, our scaling does not work well for the small offsets, 
e.g., $\alpha =1$ or $2$, where 
scaling exponents there may depend on the offset strongly.

\begin{figure}
\begin{center}
\includegraphics[width=15cm]{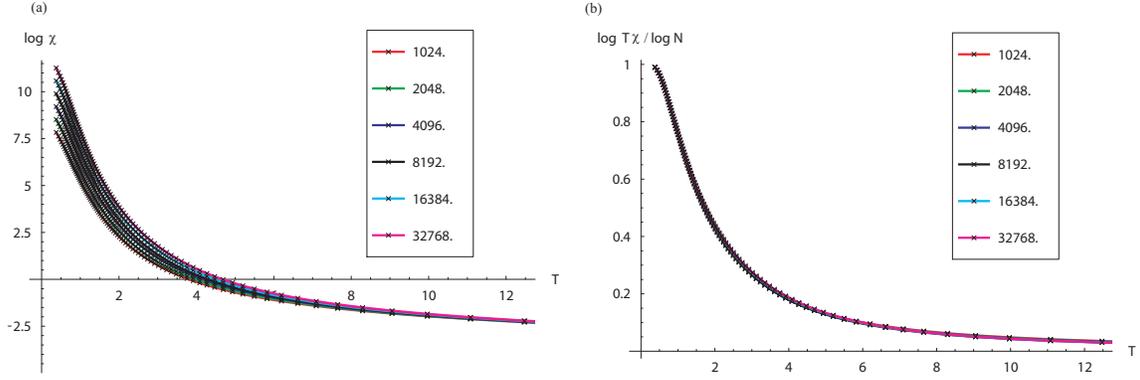}
\end{center}
\caption{
(a)The system susceptibility $\log \overline{\chi_{\rm sys}}$
and (b) $\log (T \overline{\chi_{\rm sys}}) / \log N$ 
on the GN model for the no offset $\alpha = 0$. 
The number of nodes are taken from $N=2^{10}$ to $2^{15}$.
The results are averaged over $100$ samples.
}
\label{nooffset}
\end{figure}

Finally, we consider the no offset case.
In Fig.\ref{nooffset}-(a), 
we plot the mean system susceptibilities 
$\log [\overline{\chi_{\rm sys}}(N,T)]$ 
with several nodes from $N=2^{10}$ to $2^{15}$.
Now we rescale these susceptibilities as 
$\log [T \overline{\chi_{\rm sys}}(N,T)] / \log N$. 
The rescaled system susceptibilities are quite well fitted for very wide temperature range as seen in Fig.\ref{nooffset}-(b).
This indicates that $T \overline{\chi_{\rm sys}}(N,T)$ 
goes to the infinity as $N \to \infty$. 
In addition, our result means that 
the following unusual scaling for the system susceptibility 
(in other words, for the average correlation volume $\overline{\xi_V}$) is satisfied: 
\begin{equation}
T \overline{\chi_{\rm sys}}(N,T)= [g(T)]^{\log N}, \label{scaleform}
\end{equation}
where $g(x)$ is a  scaling function. 
Unfortunately, we have not obtained the derivation of this scaling yet.
But this relation is derived partially by the following approximation.
Bollob\'as and Riordan \cite{BR04} derived that
the expected number $E_l$ of shortest paths of length $l$ 
for the Barab\'asi-Albert tree with $N$ nodes is given as 
\begin{equation}
E_l \sim \frac{N^2}{2} \frac{1}{\sqrt{2 \pi \log N}} e^{-\frac{(l-\log N)^2}{2\log N}},
\end{equation}
for $N \gg 1$.
Using this distribution, 
we approximate the system susceptibility as 
\begin{eqnarray}
T \overline{\chi_{\rm sys}}(N,T) 
&=& \frac{1}{N} \sum_{l=0}^{N-1} E_l t^l \nonumber \\
&\sim&  
\frac{N}{2} \frac{1}{\sqrt{2 \pi \log N}}
\int_0^{N-1} 
e^{-\frac{(l-\log N)^2}{2\log N}} t^l {\rm d}l \nonumber \\
&\sim& 
\frac{1}{2} e^{\log N[1+\log t+ \frac{1}{2}(\log t)^2]}, \label{BATapprox}
\end{eqnarray}
for $\log t >-1$. 
This approximation shows 
the system susceptibility $T \overline{\chi_{\rm sys}}(N,T)$
holds the scaling relation (\ref{scaleform}) at least in a low temperature region.
Interestingly, this scaling form 
remains to be satisfied even at a high temperature where 
this approximation (\ref{BATapprox}) is not valid.

\section{summary \label{summary}}

In this paper, we investigated 
the zero-field susceptibility of the Ising model 
on the GN model with the attachment kernel $A_k=k+\alpha$.
Our main finding of this paper is that the divergent temperature $T_s$ of 
the GN model with the offset $\alpha$ is given by $\tanh (J/T_s)=\alpha/2 (\alpha+1)$.
The exact expression of the susceptibility for the infinite offset, and 
the finite size scaling of the susceptibilities for the finite offsets support our estimate is exact.
The finite size scaling form (\ref{infiniteFSS}) implies that 
$\log N$ can be regarded as the characteristic system length $L$, 
so that (\ref{infiniteFSS}) can be read as 
$\overline{\chi_{\rm sys}} \sim L^\gamma f(\Delta_T L^{\frac{1}{\nu}})$ 
with $\gamma =1$, $\nu =1$.

\begin{figure}
\begin{center}
\includegraphics[width=10cm]{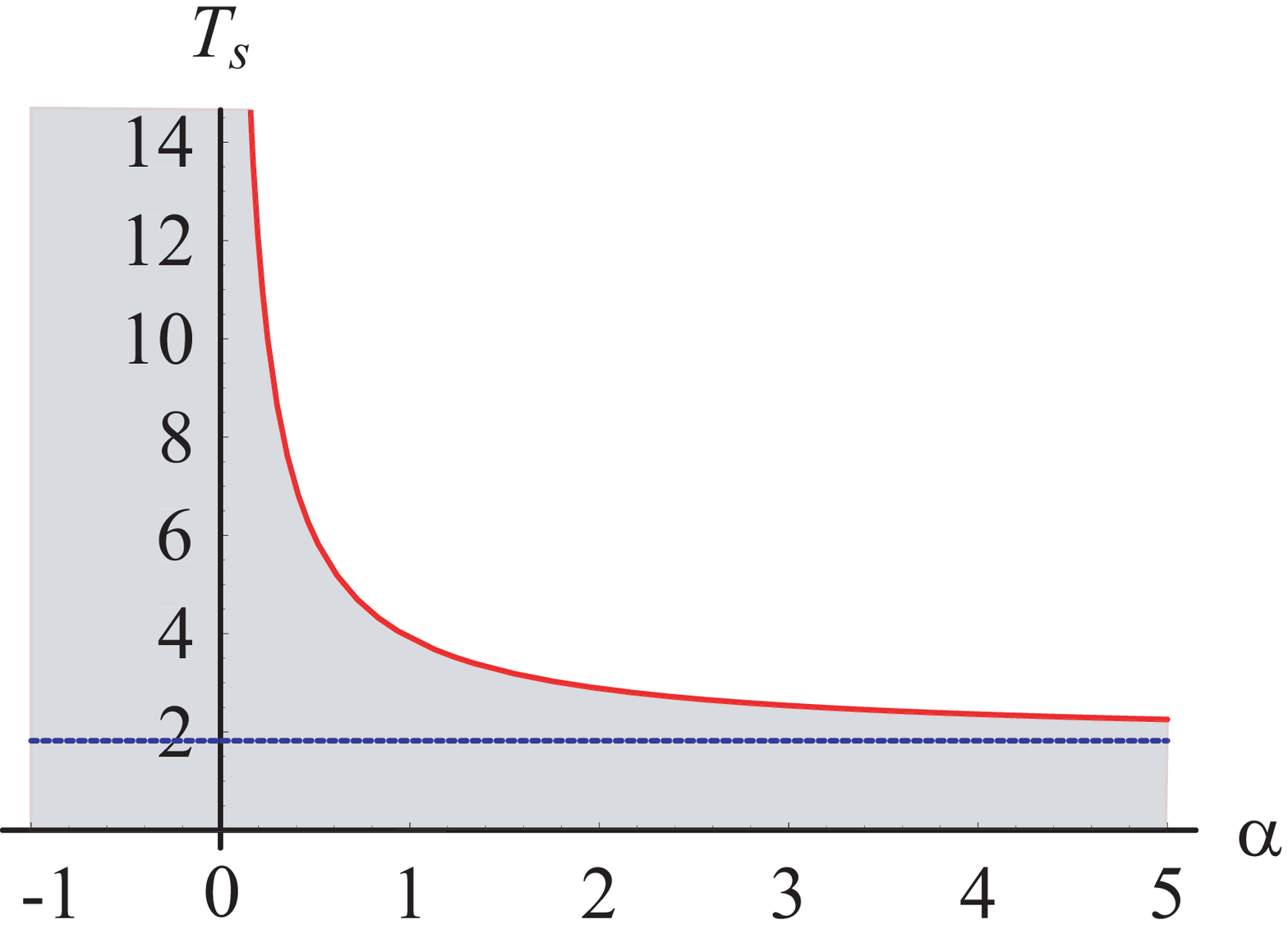}
\end{center}
\caption{
The relation between the divergent temperature $T_s$ and 
the offset $\alpha$ of the Ising model on the GN model.
The red-solid line denotes the relation (\ref{estimate}) between $\alpha$ and $T_s$.
The blue-dot line denotes $T_s$ for the infinite offset $\alpha$.
In the blue-colored region, the system susceptibility diverges.
}
\label{schemaplot}
\end{figure}

Our result means that 
as the offset $\alpha$ is smaller, the divergent temperature $T_s$ is higher (Fig.\ref{schemaplot}).
Particularly, $T_s$ diverges 
when $\alpha \le 0$, that is, the degree exponent $\gamma \le 3$.
As is pointed out in \cite{DorogoRev}, 
a long-ranged spin correlation covers the whole system below $T_s$ 
if we apply a small local external field on the node $i$.
So one finds that applying a small local field, 
or maybe adding a few shortcuts 
to spin systems on a tree with a fat-tailed degree distribution
induces a long-ranged ordering {\it at any finite temperature}.
Interestingly, our result shows that 
the susceptibilities for the no offset obey unusual scaling (\ref{scaleform}).
We will investigate the origin of this feature in the future.

\section*{Acknowledgment}

This work is supported by the 21st Century Center of Excellence (COE) program entitled "Topological Science and Technology", Hokkaido University.

\appendix
\section{the derivation of the divergence temperature of the GN model with the infinite offset \label{derdivExpoTree}}

In this appendix, we derive the divergence temperature of 
of the GN model with the infinite offset $\alpha \to \infty$.
We rewrite the system susceptibility (\ref{totalchiform}) as follows:
\begin{equation}
v_N=1+t+2t \sum_{m=2}^{N-1}\frac{1}{m(m+1)} Q_m, \label{fnandsm}
\end{equation}
where
\begin{equation}
Q_m=\prod_{k=1}^{m-1}\Big( 1+\frac{2t}{k}\Big). \label{Smrec}
\end{equation}
Now, we can derive that the divergence occurs at $t=1/2$.
At $t=1/2$, one find 
\begin{equation}
Q_m=\prod_{k=1}^{m-1}\frac{k+1}{k}=m,
\end{equation}
to reduce the system susceptibility (\ref{fnandsm}) to
\begin{equation}
v_N=1+t+2t \sum_{m=2}^{N-1} \frac{1}{m+1}=\sum_{m=1}^{N}\frac{1}{m} ,
\end{equation}
i.e., to the harmonic series.
Thus, the system susceptibility diverges at $t=1/2$ in the limit $N \to \infty$.
We find immediately that  
the system susceptibility diverges at least for $t \ge 1/2$, 
since $v_N$ increases monotonously with $t$ for any $N$. 
Now we show that the system susceptibility cannot diverge for $t <1/2$.
From Eq.(\ref{Smrec}), we obtain the following relations for $Q_m$:
\begin{equation}
\sum_{m=2}^{N-1} \frac{1}{m(m+1)} Q_m= \sum_{m=2}^{N-1} \Big( \frac{1}{m}-\frac{1}{m+1} \Big) Q_m, \label{Srelation1}
\end{equation}
and
\begin{equation}
Q_{m+1}-Q_{m}=\frac{2t}{m}Q_m. \label{Srelation2}
\end{equation}
The iterative substitutions of Eqs.(\ref{Srelation1}) and (\ref{Srelation2}) 
allow one to the following transformation:
\begin{eqnarray}
&&\sum_{m=2}^{N-1} \frac{1}{m(m+1)}Q_m \nonumber \\ 
&=& \frac{1}{2}Q_2+\sum_{m=2}^{N-2}\frac{2t}{m(m+1)}Q_m-\frac{1}{N}Q_{N-1}  \nonumber \\
&=& \frac{1+2t}{2}Q_2+\sum_{m=2}^{N-3}\frac{(2t)^2}{m(m+1)}Q_m
-\frac{2t}{N-1}Q_{N-2}-\frac{1}{N}Q_{N-1} \nonumber \\ 
&=& 
\cdots = \frac{1}{2}\sum_{m=2}^{N-2}(2t)^{m-1} -\sum_{m=2}^{N-1}\frac{(2t)^{N-m-1}}{m+1}Q_m.
\end{eqnarray}
In the end, one find 
\begin{eqnarray}
v_N = 1+t+\frac{1}{2} \sum_{m=1}^{N-2} (2t)^{m-1}- \sum_{m=2}^{N-1} \frac{(2t)^{N-1-m}}{m+1}Q_m 
\le 1+t+\frac{1}{2} \sum_{m=1}^{N-2} (2t)^{m-1}.
\end{eqnarray}
This upper bound converges for $t<1/2$, 
so $v_N$ doesn't diverge there.
Thus, we find that the divergent temperature $T_s$ is decided by 
$\tanh (J/ T_s)=1/2$.

\clearpage


\begin{thebibliography}{99} 

\bibitem{BA}A.-L. Barab\'asi and R. Albert: Science {\bf286} (1999) 509.
\bibitem{AB}R. Albert and A.-L. Barab\'asi: Rev. Mod. Phys. {\bf74} (2002) 47.
\bibitem{NewRev}M. E. J. Newman: SIAM Review {\bf 45} (2003) 167.
\bibitem{Get} K. I. Goh, E. S. Oh, H. Jeong, B. Kahng, and D. Kim: Proc. Natl. Acad. Sci. U.S.A. {\bf 99} (2002) 12583.
\bibitem{DorogoRev}S. N. Dorogovtsev, A. V. Goltsev, J. F. F. Mendes: cond-mat/0705.0010v6
\bibitem{Bi}G. Bianconi: Phys. Lett. A {\bf303} (2002) 166.
\bibitem{Let}M. Leone, A. V\'azquez, A. Vespignani, and R. Zecchina: Eur. Phys. J. B {\bf 28}(2002) 191. 
\bibitem{IT}F. Igl\'oi and L. Turban: Phys. Rev. E  {\bf 66} (2002) 036140 .
\bibitem{Det02}S. N. Dorogovtsev, A. V. Goltsev, J. F. F. Mendes: Phys. Rev. E {\bf66} (2002) 016104.
\bibitem{SH}K. Suchecki and J. A. Holyst: Phys. Rev. E {\bf 74} (2006) 011122. 
\bibitem{HB}M. Hinczewski and A. N. Berker: Phys. Rev. E {\bf 73} (2006) 066126.
\bibitem{AHS}A. Aleksiejuk, J. A. Holyst, D. Stauffer: Physica. A {\bf310} (2002) 260.
\bibitem{H}C. P. Herrero: Phys. Rev. E {\bf69} (2004) 067109.
\bibitem{ZL}H. Zhou and R. Lipowsky: PNAS {\bf 102} (2005) 10052.
\bibitem{E}T. P. Eggarter: Phys. Rev. B {\bf 9} (1974) 2989.
\bibitem{MHZ}E. M\"uller-Hartmann and J. Zittartz: Phys. Rev. Lett. {\bf 33} (1974) 893.
\bibitem{M74}H. Matsuda: Prog. Theor. Phys. {\bf 51} (1974) 1053.
\bibitem{HT} J. von Heimburg and H. Thomas: J. Phys. C {\bf 7} (1974) 3433.
\bibitem{F}H. Falk: Phys. Rev. B {\bf 12} (1975) 5184.
\bibitem{Met}R. M\'elin, J. C. Angl\`es d'Auriac, P. Chandra and B. Dou\c{c}ot: J. Phys. A: Math. Gen. {\bf 29} (1996) 5773.
\bibitem{Set98}T. Sto\v{s}i\'c, B.D. Sto\v{s}i\'c, I.P. Fittipaldi: J. Mag. Mag. Mater. {\bf185} (1998) 177; B. D. Sto\v{s}i\'c, T. Sto\v{s}i\'c, I. P. Fittipaldi: Physica A {\bf355} (2005) 346.
\bibitem{HN}T. Hasegawa and K. Nemoto: Phys. Rev. E {\bf 75} (2007) 026105; 
T. Hasegawa and K. Nemoto: Physica A {\bf 387} (2008) 1404.
\bibitem{KR00}P. L. Krapivsky, S. Redner and F. Leyvraz: Phys. Rev. Lett. {\bf 85} (2000) 4629-4632; 
P. L. Krapivsky and S. Redner: Phys. Rev. E {\bf 63} (2001) 066123; 
P. L. Krapivsky and S. Redner: J. Phys. A: Math. Gen. {\bf 35} (2002) 9517; 
P. L. Krapivsky and S. Redner: Phys. Rev. Lett. {\bf 89} (2002) 258703.
\bibitem{BR04}B. Bollob\'as and O. Riordan: Phys. Rev. E {\bf }69 (2004) 036114.
\bibitem{FVK06}A. Fekete, G. Vattay, and L. Kocarev: Phys. Rev. E {\bf 73} (2006) 046102.
\bibitem{SAK02}G. Szab\'o, M. Alava, and J. Kert\'esz: Phys. Rev. E {\bf 66} (2002) 026101.
\bibitem{MetDis}
K. Malarz, J. Czaplicki, B. Kawecka-Magiera, and K. Kulakowski: Int. J. Mod. Phys. C {\bf 14} (2003) 1201; 
K. Malarz, J. Karpi\'nska, A. Kardas, K. Kulakowski: TASK Quarterly {\bf 8} (2004) 115; 
K. Malarz, K. Kulakowski: Eur. Phys. J. B {\bf 41} (2004) 333; 
K. Malarz, K. Kulakowski: Acta Phys. Pol. B {\bf 36} (2005) 2523; 
K. Malarz: Acta Phys. Pol. B {\bf 37}(2006) 309.
\end{thebibliography}
\end{document}